# Plasmonic Nanolasers Without Cavity, Threshold and Diffraction Limit using Stopped Light

**Kosmas L. Tsakmakidis, Joachim M. Hamm, Tim W. Pickering and Ortwin Hess**
*Blackett Laboratory, Department of Physics, Imperial College London, London SW7 2AZ, United Kingdom*
*Author e-mail address: k.tsakmakidis@imperial.ac.uk; o.hess@imperial.ac.uk*

**Abstract:** We present a plasmonic waveguide where light pulses are stopped at well-accessed complex-frequency zero-group-velocity points. Introducing gain at such points results in cavity-free, "thresholdless" nanolasers beating the diffraction limit via a novel, stopped-light mode-locking mechanism.

## 1. Introduction

Reducing the velocity of light down to zero [1] is of fundamental scientific interest that could usher in a host of important photonic applications, some of which are hitherto fundamentally inaccessible. These include cavity-free, low-threshold nanolasers, novel solar-cell designs for efficient harvesting of light [2], nanoscale quantum information processing [3], as well as enhanced biomolecular sensing. Until now, complete stopping of light pulses, leading to their localisation in a specific region of space, in *solid* structures and at *ambient* conditions, has been hampered by fundamental difficulties. Ultraslow light requires strong group-index ($n_g$) resonances but the increased (compared with gases) damping at normal conditions invariably broadens and weakens such resonances.

We report on a solid-state configuration leveraging low-loss plasmonic media (indium tin oxide, ITO), whereby an arbitrary number of light pulses can be stopped, remaining well localised – without diffusing – at pre-defined spatial locations despite the absence of any barriers in front or behind the pulses. On the basis of a full-wave Maxwell-Bloch approach [4-6], we show how by introducing gain in a small region inside this structure we can construct a cavity-free nanolaser that does not have a threshold and operates well below the diffraction limit set by the wavelength of the emitted (coherent) light inside the structure.

## 2. Complex-$\omega$ stopping and localization of light in plasmonic heterostructures

We have considered two light-stopping configurations. First, a microwave negative-index slab surrounded by air, as shown in Fig. 1(a) (not detailed further herein), and second an ITO/Si/ITO heterostructure, illustrated in Figs. 1(b)-(d) below. For the permittivity of ITO we used the experimental Drude parameters of Ref. [7]. We calculated the complex dispersion bands of the *photonic* modes of the plasmonic structure, and identified a zero-

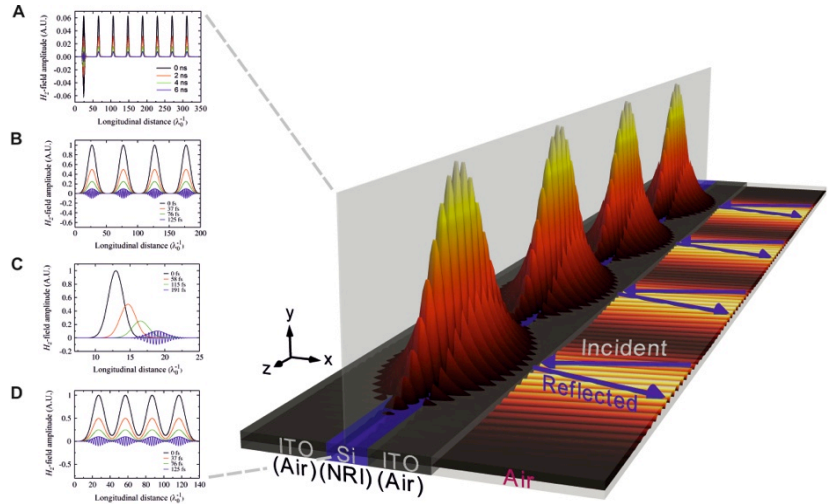

Fig. 1. Full-wave FDTD calculations of the time-evolution of one or more pulses injected inside a plasmonic (ITO) or negative-index waveguiding heterostructure. In all cases, the results are recorded on the vertical *yz* plane (translucent) cutting through the middle layer. (**a**) Shown are eight pulses in-coupled into the negative-index heterostructure at its zero-$v_g$ point. All pulses remain stopped for more than 6 ns, decaying with time at their initial positions. (**b**) Shown here are four pulses in-coupled into the plasmonic heterostructure at its at its zero-$v_g$ point. The pulses remain stopped and localized, without broadening, for more than 130 fs. (**c**) When the evanescent in-coupling is performed at a non-zero-$v_g$ point, a pulse in-coupled into the plasmonic structure moves away from its initial point, both decaying and broadening with time. (**d**) The pulses in **b** can also be brought closer together – up to the diffraction limit – since the structure is completely uniform, and the localization is not aided by disorder or a cavity-like action.

$v_g$ point for the supported complex-frequency [8] mode (Fig. 2(b)) – resilient to realistic levels of dissipative, radiative and surface-roughness losses. At that point, the band is flat (group-velocity dispersion is minimised) and the complex-$\omega$ mode is weakly leaky; hence it can be accessed via evanescent in-coupling directly from free space without a need for a prism or a grating to add momentum (see Fig. 1). When the incidence angle of the coupling light beams is chosen such that the zero-$v_g$ point is precisely hit, all injected light pulses remain stopped and localised (Fig. 1(b)), without diffusing/broadening, at predefined spatial locations along the heterostructure. We shall show that although the so-attained group-refractive indices are extraordinarily high for a solid-state structure ($n_g \sim 10^6$), this is achieved with only a moderate increase of the optical losses (by a factor of approximately 1.5). The lifetimes of the stopped and localized (absence of diffusion) light pulses are typical nanoplasmonic ones, thereby potentially allowing for a host of applications in this field.

### 3. Plasmonic nanolasers at the stopped-light regime

Next, we introduce gain in a small region of the stopped-light waveguide, as shown in Fig. 2(a). In our Maxwell-Bloch (MB) analysis of this active nanoplasmonic configuration the gain is modelled as a four-level medium [4-6]. The emission frequency was chosen to correspond exactly to the zero-$v_g$ point, whereas pumping was performed at a higher frequency (where the group velocity is non-zero and the pump field can propagate along the guide) – see Fig. 2(b). The difference in the real part of the refractive index of the gain medium with that of the adjacent dielectric (in the longitudinal direction) is negligibly small, of the order of $\sim 10^{-3}$, i.e. there is no Fabry-Perot cavity formed in the structure.

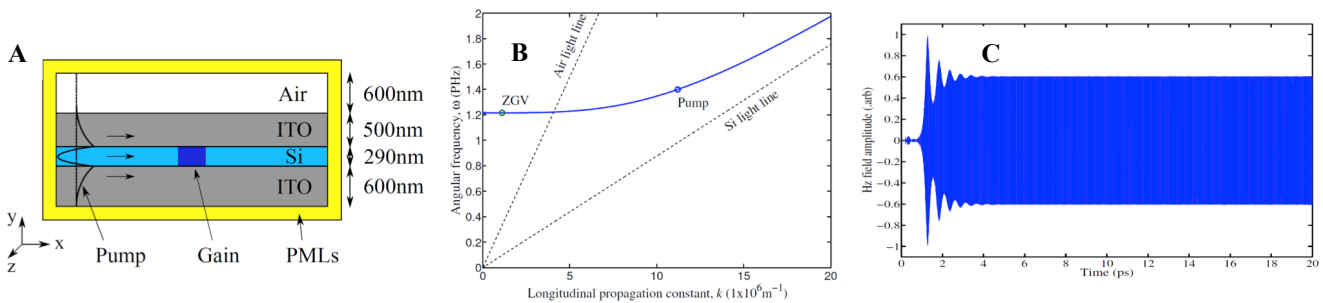

Fig. 2. (**a**) Schematic illustration of the computational setup for the modelling of the stopped-light plasmonic nanolaser. (**b**) Dispersion diagram of the complex-frequency photonic (i.e., not surface-wave) mode, showing the location of the zero-group-velocity (ZGV) point. The emission frequency of the gain medium exactly coincides with the ZGV one, whereas pumping is performed at a higher frequency. Note that the ZGV point is on the left-hand side of the air light line, i.e. the mode at that point is (weakly) leaky. (**c**) Lasing relaxation oscillations and steady-state quasi-monochromatic emission from the cavity-less plasmonic nanolaser, emitting at the stopped-light frequency.

We shall show that by centring the emission spectrum of the gain medium around the ZGV point, a broad range of longitudinal wavevectors (in the region where the band is flat, around the ZGV point in Fig. 2(b)) are pulled towards the most stable (maximum gain) ZGV point, giving rise to a deep-subwavelength localisation (the lasing emission frequency is much larger than the gain-medium dimension in the longitudinal direction), *without having to use surface or evanescent waves*. This mechanism is similar to the creation of ultrashort light pulses in mode-locked lasers, but here it arises in the absence of a cavity, solely by accessing a light-stopping point. Furthermore, because at the ZGV point the laser $\beta$-factor tends to unity ($\beta \to 1$), spontaneous emission is efficiently funnelled into the stopped-light mode, resulting in "thresholdless" operation [9]. Figure 2(c) reports an exemplary result of full-wave MB calculations showing how, after pumping of the gain medium, a transient phase ensues characterized by sup-ps relaxation oscillations, followed by coherent, quasi-monochromatic emission. We shall present a range of further results demonstrating the attainment of cavity-free, "thresholdless" lasing operation, deep below the diffraction limit set by the wavelength (1.55 μm) of the emitted light.

### 4. References

[1] K. L. Tsakmakidis, A. D. Boardman and O. Hess, "'Trapped rainbow' storage of light in metamaterials," *Nature* **450**, 397-401 (2007).
[2] A. Aubry, D. Yuan, A. I. Fernández-Domínguez, Y. Sonnefraud, S. A. Maier and J. B. Pendry, "Plasmonic light-harvesting devices over the whole visible spectrum," *Nano Lett.* **10**, 2574-2579 (2010).
[3] Z. Jacob and V. M. Shalaev, "Plasmonics goes quantum," *Science* **334**, 463-464 (2011).
[4] O. Hess, J. B. Pendry, S. M. Maier, R. F. Oulton, J. M. Hamm and K. L. Tsakmakidis, "Active nanoplasmonic metamaterials," *Nature Materials* (review), in press (2012).
[5] S. Wuestner, A. Pusch, K. L. Tsakmakidis, J. M. Hamm and O. Hess, "Gain and plasmon dynamics in active negative-index metamaterials," *Phil. Trans. R. Soc. A* **369**, 3525-3550 (2011).
[6] A. Pusch, S. Wuestner, J. M. Hamm, K. L. Tsakmakidis and O. Hess, "Coherent amplification and noise in gain-enhanced nanoplasmonic metamaterials: A Maxwell-Bloch Langevin approach," *ACS Nano* **DOI:** 10.2021/nn204692x (2012).
[7] M. A. Noginov *et al*., Transparent conducting oxides: Plasmonic materials for telecom wavelengths. *Appl. Phys. Lett.* **99**, 021101 (2011).
[8] E. I. Kirby, J. M. Hamm, T. W. Pickering, K. L. Tsakmakidis and O. Hess, "Evanescent gain for slow and stopped light in negative refractive index heterostructures," *Phys. Rev. B* **84**, 041103(R) (2011).
[9] S. Noda, "Seeking the ultimate nanolaser," *Science* **314**, 260-261 (2006).